\documentstyle[prb,aps,epsfig,twocolumn]{revtex}

\begin{document}

\draft

\title{Vortices in small superconducting disks}
\author{E.~Akkermans and K.~Mallick\cite{kirone}\\ Department of Physics,
 Technion, 32000 Haifa, Israel}

\maketitle

\begin{abstract}
 We  study   the Ginzburg-Landau equations in order to describe
a two-dimensional superconductor in a bounded domain. Using the properties of 
a particular integrability point ($\kappa = 1/ \sqrt2$) of these nonlinear 
equations which allows vortex solutions,
 we obtain a closed expression for the energy of the superconductor. The 
presence of the 
boundary provides a selection mechanism for the number
 of vortices.

 A  perturbation analysis
 around $\kappa = 1/ \sqrt2$ enables us  to include the effects of the 
vortex interactions and to describe quantitatively
 the magnetization curves recently measured on small
 superconducting disks \cite{geim}.
 We also  calculate  the optimal vortex configuration
  and obtain an expression for the confining potential 
 away from the London limit.  
\end{abstract}
 

\def\real{{\rm I\kern-.2em R}}
\def\complex{\kern.1em{\raise.47ex\hbox{
	    $\scriptscriptstyle |$}}\kern-.40em{\rm C}}
\def\integer{{\rm Z\kern-.32em Z}}
\def\pinteger{{\rm I\kern-.15em N}}

The dimensionless Ginzburg-Landau energy functional $\cal F$ 
of a superconductor depends on only one parameter \cite{degennes},
 the ratio $\kappa$ between the London penetration depth 
 $\lambda$ and the coherence length $\xi$ 
\begin{equation}
{\cal F} = {\int_{\Omega}} {1 \over 2} |B{|^2} + {\kappa^2}
 |1 - |\psi{|^2}{|^2} + 
|({\vec{\nabla}} - i{\vec A})\psi {|^2} .
\label{energy}
\end{equation}
The  order parameter $\psi$ is dimensionless 
as well as the magnetic field $B$
 measured in units 
of ${ {\phi_0} \over {4 \pi \lambda^2}}$, with
 ${\phi_0} = { hc \over 2e}$. 
Lengths are measured  in units of 
$\lambda { \sqrt 2} $. The  expression (\ref{energy})
 assumes that both the order 
parameter and the vector potential have
 a slow spatial variation. The integral is over the volume 
$\Omega = \pi R^2 d$ of a thin disk of radius $R$ and thickness $d$.

Outside the superconducting sample, the order
 parameter vanishes and the magnetic field 
is solution of the Maxwell equation. The boundary condition
 between a superconductor and an insulator is 
$(\vec \nabla - i \vec A ) 
\psi{|_{\vec n}} =0 $ 
where $\vec n$ is the unit vector normal to the surface of the disk.
 The presence of a boundary precludes a complete analytical solution 
of the 3d Ginzburg-Landau equations for a thin disk. 
We are then led to make some simplifying  assumptions,
 based upon numerical results \cite{deo}.

The thickness $d$ of the sample considered in the
 experiments \cite{geim} fulfills 
$d \ll \xi$ and $d \leq \lambda$. 
If the curvature of the magnetic flux lines, given by $R / \lambda_e^2 $ 
(where $\lambda_e ( d, R, \lambda)$ stands for
the effective screening length),  is 
smaller than $1 / \lambda_e $, {\it i.e.} if  $ R \ll \lambda_e $, then both 
$\psi$ and the vector potential $\vec A$ can be considered to be
 constant accross 
the thickness and   the disk is effectively two-dimensional.
 The expression of the effective screening length $\lambda_e (d, R, 
\lambda)$ is not known, 
except for  the   case $R \rightarrow \infty$  where
\cite{pearl,fetter}  $\lambda_e \simeq \lambda^2 / d$.
 In the London  limit ({\it i.e.} $\kappa \to \infty$), such a 
  system has  been  described using 
Pearl's  solution \cite{pearl}. 
Finally, 
since $\psi$ and $\vec A$ are constant over the thickness,
 the covariant Neumann  boundary condition, stated  above,  
is automatically satisfied  on the upper and lower surface of the disk. 

  The Ginzburg-Landau equations  are 
nonlinear, second order differential equations 
 whose solutions are usually unknown.
 However,  for the special value $\kappa = 
{1 \over \sqrt 2}$ known as the dual point \cite{bogo}, the 
equations for $\psi$ and ${\vec A}$  reduce to  first 
order differential equations and the minimal
 free energy can be  calculated exactly
for an infinite plane.
This relies on the identity true for two dimensional systems
$ |({\vec{\nabla}} - i{\vec A})\psi {|^2}= |{\cal D} \psi {|^2} +
 {\vec \nabla} \times {\vec \jmath} + B |\psi{|^2}$
where ${\vec \jmath}$
is the current density and the operator $\cal D$ is 
defined as  ${\cal D} = \partial_x + i\partial_y -i(A_x + iA_y)$. 
  At the dual  point, the
 expression (\ref{energy}) for   $\cal F$  is  rewritten using this 
identity as follows 
\begin{equation}
{\cal F} = {\int_{\Omega}} {1 \over 2} | B - 1 + |\psi{|^2}{|^2} +
  |{\cal D} \psi {|^2}
 + {\oint_{\partial \Omega}} ({\vec\jmath} + {\vec A}).{\vec dl}
 \label{identitebog}
\end{equation}where the last integral over the boundary 
${\partial \Omega}$ of the
 system results from  Stokes theorem.

 For an infinite plane, we impose that 
the system is superconducting at large distance, 
  i.e. $|\psi| \to 1$ and
 ${\vec \jmath} \to 0$  at infinity so that the boundary
 term in (\ref{identitebog}) 
coincides with the  London
fluxoid. It is quantized and equal to
 $\oint_{\partial\Omega} {\vec \nabla}\chi.{\vec dl}= 2 \pi n $,
 where $\chi$ is the phase of the order parameter.
  The integer $n$ is 
   the winding number of the order parameter $\psi$ and as such is a
 topological characteristic of the system \cite{houches}.
   The extremal values of ${\cal F}$, 
 are ${\cal F} = 2 \pi n$, and are  obtained
 when the bulk  integral in (\ref{identitebog})
 vanishes identically giving rise to two first order 
differential equations.
 These two equations can be decoupled to give for 
   $|\psi|$ a second order nonlinear equation 
which admits families of vortex solutions 
\cite{taubes}. However,  for the infinite plane,
 there is no mechanism to  select the value of $n$, which
 only plays the role of a classifying parameter. 
   
The extension of these results to  
finite size systems, namely the existence and stability
 of vortex solutions and 
their  behaviour  as a function of the applied field received
 a partial numerical answer. 
Numerical simulations of the Ginzburg-Landau equations \cite{argentins}
 show the existence of stationary vortex solutions
 whose number depends on the applied  magnetic field.
 Moreover, these simulations indicate that the physical picture  derived for
$\kappa = {1 \over \sqrt 2}$ 
remains qualitatively valid  for quite a large range of values of  $\kappa$,
 with a small corresponding  change of free  energy \cite{rebbi} . 
 
We  consider finite size systems at the dual point i.e.
 for $\kappa = {1 \over \sqrt 2}$.
There, the edge currents
 screen the external magnetic field therefore producing
 a magnetic moment   
opposite to the direction of the field, 
 whereas  vortices in the bulk of the system produce a
 magnetic moment along the direction of the applied field.
 Assuming cylindrical symmetry, the current density
 $\vec\jmath$   has only 
 an azimuthal component, with  opposite  signs  in the bulk
 and on the edge of  the system.
 Thus,  there exists  a circle $\Gamma$  on which  $\vec\jmath$ vanishes
 \cite{remark}.
 This allows us to separate  the domain $\Omega$ into two concentric subdomains
 $ {\Omega_1}$ and $ {\Omega_2}$ whose boundary is  the circle
 $\Gamma$. Therefore one can extend to the subdomain $\Omega_1$ the results 
obtained for the infinite case.
 The existence of vortices in a finite domain such as
   $\Omega_1$  was  checked  numerically \cite{devega}. It was shown that  $|\psi|$  vanishes as a power law 
  at  the center of the disk,
 hence  there is a (multi-)vortex in  the center
 whose multiplicity is determined by the  exponent of the power law.
  The  magnetic flux $\Phi ({\Omega_1}) = n$
 is quantized and the free energy in $\Omega_1$  is 
$ {\cal F}({\Omega_1})  = 2 \pi n $.

  The contribution of ${\cal F}(\Omega_2)$
 to the free energy  can be written,  using
the phase and the modulus of the order parameter $\psi$, as
 \begin{equation}
   {\int_{\Omega_2}}  (\nabla |\psi|{)^2}
 + |\psi({\vec{\nabla}}\chi - {\vec A})|^2
  + \frac { B^2 +  (1 - |\psi|^2)^2 } {2}
\label{tran1}
\end{equation}
  We know, from  the London equation, that both 
the magnetic field and the vector potential
  decrease rapidly  away from
 the boundary $\partial\Omega$ of the system over a
 distance of order $\lambda {\sqrt2}$.
  Over the same distance, at the dual point,   $|\psi|$ saturates to unity. 
  One can  thus estimate the integral (\ref{tran1})
 using a saddle-point method. 
 We assume  cylindrical symmetry, and we  neglect 
 the term $(\nabla |\psi|{)^2}$  on the boundary  because of 
 the  boundary conditions, so that the relation (\ref{tran1}) is now 
 given by an integral over the boundary of the system. 
 To go further, we need to implement
 boundary conditions for  the magnetic field $ B(R)$ 
 and the vector potential $A(R)$.
 The  choice  $ B(R) =  {B_e}$, where $B_e$ is the external imposed field,
  corresponds  to the geometry of an infinitely long cylinder,
 where the flux lines are not distorted outside the system.
 A more suitable choice for a flat thin  disk is provided by demanding
  $\phi = {\phi_e}.$ 
 This boundary condition implies that
  the vector potential is  identified by continuity
 to its external applied  value ${\vec {A_e}}$.
 It should be noticed that
  the magnetic field ${\vec B}$ has then a non monotonous  variation:
  it is low in the bulk, larger than $B_e$ near the edge of the system,
 because of the distortion of flux lines,
 and eventually equal to  its applied value 
 far  outside the system \cite{deopeeters}.

 Finally, the minimization of the free energy with respect to $|\psi|$ gives
 $1 - |\psi{|^2} = |{\vec \nabla} \chi - {\vec A}{|^2}$,
 such that, performing the integral over the boundary of the system, we 
obtain 
\begin{equation}
{1 \over {2 \pi}}{\cal F}(\Omega_2) = {{ \lambda {\sqrt 2}} \over R}
 (n - {\phi_e} {)^2}
- {1 \over 2} ({{\lambda {\sqrt 2}} \over R}{)^3}(n - {\phi_e} {)^4}
\label{free2}
\end{equation}
 We have neglected the contribution of the  $B^2$ term,
 which is smaller  by a factor
 of the order $(\lambda/R)^2$.

 The thermodynamic Gibbs potential ${\cal G}$ of the system is then 
\begin{equation}
{1 \over {2 \pi}} {\cal G}(n, {\phi_e}) = n + a
 (n - {\phi_e} {)^2}
- { a^3 \over 2} (n - {\phi_e} {)^4}
 - a^2 {\phi_e}^2
\label{gibbs}
\end{equation}
 where we have defined $a = {{ \lambda {\sqrt 2}} \over R}$.
The  relation (\ref{gibbs})  consists in a set of quartic
 functions indexed by the integer $n$. The minimum 
of the Gibbs potential is the envelop curve defined by the equation 
${{\partial {\cal G}} \over {\partial n}}{|_{\phi_e}} =0$, i.e.
 the system chooses its winding number 
$n$ in order to minimize $\cal G$. This provides a relation between 
the number $n$ of vortices in the system and the applied magnetic field 
$\phi_e$. 

We consider  the limit of  large enough 
$ R \over \lambda$, such that the quartic
 term is negligible. The 
Gibbs potential then reduces to a set of parabolas.
 The vortex number $n$ is then given by the integer part 
\begin{equation}
n = [  { \phi_e}
 - {R \over {2 {\sqrt2}\lambda}} + {1 \over 2}  ] 
\end{equation}
while the magnetization
 $ M= - {  {\partial {\cal G}}   \over  {\partial \phi_e}  },$ is given by
\begin{equation}
 - M =  2 a  ( {\phi_e} - n ) -
 2 a^{2}  {\phi_e}
\label{magnet}
\end{equation}
For ${\phi_e}$ smaller that $R \over {2{{\sqrt 2}\lambda}}$,
 we have $n =0$  and 
$(- M)$  increases linearly with 
 the external flux. This corresponds 
to the London regime.
 The field $H_1$ at which the first vortex enters the disk corresponds to ${\cal G}(n=0)=
 {\cal G}(n=1)$, i.e. to 
\begin{equation}
  {H_1} = { {\phi_0} \over {2 \pi {\sqrt 2} R \lambda} }
 + { {\phi_0} \over {2 \pi  R^2} }
\end{equation}
 The subsequent vortices enter one by one for each crossing ${\cal G}(n+1)= {\cal G}(n)$; 
 this happens periodically in the applied field, with a period equal to
${\Delta}H = {{\phi_0} \over { \pi  {R^2}}}$ and a discontinuity of the magnetization
 $ {\Delta}M = {{2 {\sqrt 2} \lambda} \over R }.$

There is a qualitative similarity between the results
 we derived using the properties of the dual point and those obtained
 from a linearised version of the 
Ginzburg-Landau functional \cite{zwerger}. However, 
 the two approaches differ in 
their quantitative predictions due to the importance of the nonlinear term.

  Within the previous approximations,  
  the  expression (\ref{gibbs}) captures
   the main  features observed experimentally i.e. the
 behaviour of the 
magnetization at low fields (before the first discontinuity), the
 periodicity and the 
linear behaviour between the successive jumps. 
From the  experimental 
parameters \cite{geim}
 namely $ R = 1.2 {\mu}m $ and 
 $ \lambda (T) = 84 nm$ 
 at $T = 0.4 K$, we compute from our expressions $ {H_1} = 25 G $ and 
 $ {\Delta}H = 4.6 G .$ These 
 values agree with the experimental results \cite{deo} 
 to within a few percent. We emphasize 
 that $ H_1$ scales like  $1 \over R$,
 whereas $ {\Delta}H$ scales like $ 1 \over {R^2}$ in accordance with the
 experimental data  \cite{geim}. We calculate the ratio of the magnetization
 jumps to the maximum value of $M$ to be 
 0.20 as compared to a measured value of 0.22. The total
 number of jumps scales like 
 ${R^2}$ and the upper critical field is independent of $R$ in our 
theory in agreement with the experimental data.

At the duality point $ \kappa = 1/ \sqrt 2$,
 the contribution of the vortices 
to the free energy is topological and does not
 depend neither on the precise shape 
of the vortices nor on the form of their interaction. 
This property does not hold for other values 
of $\kappa$. For a 2d film with an infinitesimal current sheet,
 the vortex configuration 
has been computed by Pearl \cite{pearl},
 using the London equation,  and differs qualitatively
 from the present model.
Indeed,  away from the dual point, both the 
shape of the vortices and their interaction modify 
the free energy and the magnetization.
 For instance, in the London limit ($\kappa \rightarrow 
\infty$), radially symmetric solutions become unstable
 and different geometrical configurations 
of the vortices have different energies. This results from two
 contributions to the energy, 
arising  from the interaction between
 the vortices themselves and between the vortices and the edge 
currents. We have performed  a perturbative analysis around the dual point,
 and  we obtained  that the bulk free energy ${\cal F} (\Omega_1)$
 is given by
\begin{equation}
{1 \over 2 \pi}{\cal F} (\Omega_1) = n (1 + {1 \over 2}(\kappa {\sqrt 2} -1))
 + \beta (\kappa {\sqrt 2} -1) \sum_{i < j} 
{\cal U} (r_{ij}) .
\end{equation}
The part which is  linear in $n$ is independent
 of the position of the vortices. It  has been evaluated using  a variational 
ansatz \cite{rebbi}.  The two-body interaction potential near 
   the dual point 
 is well approximated by a function ${\cal U} (r_{ij}) = {\cal U} (r) $ where 
${\cal U} (0) = 1$, ${\cal U} (\infty) = 0 $ with  
$\beta = {1 \over 4}$. In particular, 
 for a configuration where all the vortices
 are close to the center of the disk,  the bulk free energy is
\begin{equation}
{1 \over 2 \pi} {\cal F} (\Omega_1) = n (\frac{5}{8} + \kappa \frac{3 \sqrt 2}{8})
 + {1 \over 8} n^2 (\kappa {\sqrt 2} -1) 
\end{equation}
 Thus,  away from the dual point, the linear term in the bulk free energy
 is not topological anymore and is 
modified by the interaction. The attractive or repulsive character of the 
interaction between vortices depends on the sign of 
$(\kappa {\sqrt 2} -1)$.  

To obtain the edge contribution to the energy,
 we consider first the case of a single vortex placed at a 
distance $x$ (in units of $\lambda \sqrt 2$) from the center of the disk. Then, the phase of the order 
parameter is given by \cite{am2}
 $\mbox{tan} \chi = {\mbox{sin} \theta \over \mbox{cos} 
\theta - y}$, where $y = a x = x/R$  (whereas one has  
$\chi = n \theta$ for the symetrical case). 
Starting from the expression (\ref{tran1}), we obtain
\begin{equation}
{1 \over 2 \pi} {\cal F}(\Omega_2, y) = a (n - \phi_e)^2 - 
{a^3 \over 4 \kappa^2} (n - \phi_e)^4 + f(a,y, \phi_e)
\label{bl}
\end{equation}
where the function $ f(a,y, \phi_e)$
 can explicitely be calculated and represents a 
 vortex confining potential inside a finite
 superconductor  for $\kappa \simeq 1/ \sqrt 2$. This corresponds
  to the  well-known 
Bean-Livingston confining energy barrier in a 3d superconductor
 which has been  obtained in 
the extreme type II limit  \cite{bl}  using the London  equation.
It is important to emphasize that around the dual point,
 vortices are not point-like and therefore the usual 
  expression of the Bean-Livingston energy barrier does not hold.

The possible equilibrium configurations of vortices result 
from the competition between the bulk and 
edge contributions to the free energy derived above.
 It is either a giant vortex at the center of the disk, a situation 
which preserves the cylindrical symmetry, or a polygonal pattern of 
small vortices. In order to evaluate the energy of these
 configurations, we generalize the relation  
(\ref{bl}) to the case of a polygonal configuration of
 vortices placed at a distance $x$ from the 
center of the disk. The resulting energy \cite{am2}
 is $n$-times the barrier contribution $f(a,y, \phi_e)$ obtained 
in (\ref{bl}) provided the following substitutions are made: $a \rightarrow na$, $y \rightarrow y^n$ and 
$\phi_e \rightarrow {\phi_e \over n}$. 

In conclusion, we have investigated the question
 of the existence and stability of vortices in small 
two-dimensional bounded superconducting systems. 
 We have shown \cite{am1} that starting from the 
exact solution of the Ginzburg-Landau equations for an infinite plane and
 for the special value $\kappa = 1 / \sqrt 2$,
 it is possible to derive an analytical expression for the free 
energy in a bounded system. The resulting 
expression provides a satisfactory quantitative description 
of the magnetization measured on small superconducting 
aluminium disks in the low magnetic field regime. 
For larger fields, we cannot neglect anymore the interaction
 effects due to the vortices and the 
edge currents. Perturbation theory \cite{am2} around
 the value $\kappa = 1 / \sqrt 2$, has allowed us to derive  
an expression for both the confining potential barrier
 of the vortices and the strength of the interaction 
between vortices. This provides a more refined
 description of the measured magnetization at larger fields.

{\bf Acknowledgment}
K.M. acknowledges support  by the Lady Davies foundation and 
E.A. the very kind hospitality of the Laboratoire de Physique 
des Solides and the LPTMS at the university of Paris (Orsay).


\begin{thebibliography}{article}

\bibitem[\dagger]{kirone} Permanent address: Service de
 Physique Th{\'e}orique, Centre d'Etudes Nucl{\'e}aires de Saclay, 
	91191 Gif sur Yvette cedex, France
\bibitem{geim} A.K. Geim, I.V. Grigorieva, S.V. Dubonos, J.G.S. Lok, J.C. 
 Maan, A.E. Filippov and F.M. Peeters, Nature (London), {\bf 390},259 (1997)
\bibitem{degennes}  P.G. de Gennes, "Superconductivity of metals and alloys."
 Addison-Wesley (1989)
\bibitem{peeters} V.A. Schweigert and F.M. Peeters, 
 Phys.Rev. {\bf B57}, 13817 (1998)
\bibitem{deo} P. Singha Deo, V.A. Schweigert, F.M. Peeters and A.K.
 Geim, Phys.Rev.Lett. {\bf 79},4653 (1997)
\bibitem{pearl} J. Pearl, App.Phys.Lett. {\bf 5}, 65 (1964)
\bibitem{fetter} A.L. Fetter, Phys.Rev. {\bf B22}, 
 1200 (1980) 
\bibitem{bogo} E.B. Bogomol'nyi, Sov.J.Nucl.Phys. {\bf 24},449 (1977)  
 and D. Saint-James, E.J. Thomas and G. Sarma, "Type II
 Superconductivity", Pergamon Press, (1969)
\bibitem{houches} For a discussion  of  topological aspects
 see  E. Akkermans and K. Mallick, Geometrical description
 of vortices in Ginzburg-Landau billiards, {\it to appear in }
 `Topological aspects of low dimensional systems' Les Houches
 Lecture Notes (July 1998) Cond-mat/9907441
\bibitem{taubes} C. Taubes, Comm. Math.Phys. {\bf 72}, 277 (1980)
\bibitem{argentins} C. Bolech, G.C. Buscaglia and A. Lopez, Phys. Rev.
 {\bf B 52}, R15719 (1995)
\bibitem{rebbi} L. Jacobs and C. Rebbi, Phys.Rev. {\bf B 19}, 4486 (1979)
\bibitem{am1} E. Akkermans and K. Mallick, J.Phys. {\bf A 32}, 7133  
 (1999), Cond-Mat/9812275
\bibitem{remark} This property seems to   remain true
 even in the absence of  cylindrical symmetry (17)
\bibitem{devega} H.J. de Vega and F.A. Schaposnik, Phys. Rev. {\bf D 14},
 1100 (1976)
\bibitem{deopeeters} P. Singha Deo, V.A. Schweigert and F.M. Peeters,
 Phys.Rev.Lett. {\bf 81}, (1998)
\bibitem{zwerger} R. Benoist and W. Zwerger, Z.Phys. {\bf B 103},377 (1997)
\bibitem{am2} E. Akkermans and K. Mallick, Interacting vortices in small superconducting disks,
 Technion preprint (1999)
\bibitem{bl} C.P. Bean 
and J.D. Livingston, Phys. Rev. Lett. {\bf 12},14 (1964) 
 and A.I. Buzdin and J.P. Brison, Phys. Lett. {\bf A 196},267 (1994)

\end{thebibliography}
\end{document}